\documentclass[%
 reprint,
 amsmath,amssymb,
 aps,
]{revtex4-2}

\usepackage{graphicx}
\usepackage{dcolumn}
\usepackage{bm}
\usepackage{physics}
\usepackage[dvipsnames]{xcolor}
\usepackage{float}
\usepackage{xcolor}
\usepackage[normalem]{ulem} 

\usepackage{hyperref}
\usepackage{cleveref}

\begin{document}

\preprint{APS/123-QED}

\title{Cavity-induced switching between Bell-state textures in a quantum dot}

\author{S. S. Beltrán-Romero}
\email{ss.beltran@uniandes.edu.co}
\author{F. J. Rodríguez}%
\email{frodrigu@uniandes.edu.co}
\author{L. Quiroga}
\email{lquiroga@uniandes.edu.co}
\affiliation{Department of Physics, Universidad de los Andes, A.A. 4976, Bogotá D.C, Colombia}
\author{N. F. Johnson}
\email{neiljohnson@email.gwu.edu}
\affiliation{Department of Physics, George Washington University, Washington D.C. 20052, U.S.A.}
\date{\today}

\begin{abstract}
Nanoscale quantum dots in microwave cavities can be used as a laboratory for exploring electron-electron interactions and their spin in the presence of quantized light and a magnetic field. We show how a simple theoretical model of this interplay at resonance predicts complex but measurable effects. New polariton states emerge that combine  spin, relative modes, and radiation. These states have intricate spin-space correlations and undergo polariton transitions controlled by the microwave cavity field. We uncover novel topological effects involving highly correlated spin and charge density, that display singlet-triplet and inhomogeneous Bell-state distributions. Signatures of these transitions are imprinted in the photon distribution, which will allow for optical read out protocols in future experiments and nanoscale quantum technologies.
\vskip0.1in


\end{abstract}

\maketitle


\section{Introduction}

The optical control of structural, vibrational, and transport properties in nanostructure systems has led to new research areas at the intersection of quantum optics and condensed matter \cite{De_Bernardis_2018, RevModPhys.91.025005}. It has been shown that ultra-strong coupling (USC) between light and matter is crucial for exploring quantum states of matter \cite{Frisk_Kockum_2019, RevModPhys.91.025005}, quantum phase transitions, manipulating crystal structures and symmetry-breaking \cite{appugliese2021breakdown}, obtaining tunable light-matter entangled states \cite{opticexpressfarooqui}, including Floquet-Bloch states \cite{Wang_2013}. Quantum electrodynamic microcavities in solid-state systems are widely utilized to entangle photon modes and condensed matter states, leading to the creation of emergent phases with enhanced non-local correlations and the development of complex quantum materials \cite{De_Bernardis_2018, RevModPhys.91.025005}. 
This interplay between cavities and atoms enables the control of spin states using resonant optical fields \cite{Mi_2018}. Additionally, the introduction of a magnetic field or quantized polarized light can trigger chiral and topological effects \cite{lagoudakis}, which have promising applications for magnetic data storage technologies \cite{Tiecke_2014}.

Nanostructures such as semiconductor quantum dots (QDs) represent promising solid-state platforms in a wide range of contexts, including the fundamental study of strongly correlated many-body systems \cite{PhysRevX.11.041025} and applications as ideal solid-state sources of highly coherent light \cite{PhysRevB.78.125318, 2021prb}. QDs integrated into cavity systems, have attracted a lot of attention as candidate quantum emitters.  Extensive research has also been conducted on the effects of magnetic fields on nanostructures, with a specific focus on spin-dependent interactions \cite{Quiroga1993SpatialCO, LossSO}. 
Zeeman--Spin-Orbit-Coupling (SOC) competition, which is influenced by magnetic fields, is a key factor in controlling electronic behavior \cite{spintronics, LuoNature, Chakraborty}. Additionally, topological insulator states in quantum matter have emerged as a result of SOC effects in few-particle systems \cite{RevModPhys.83.1057}.

Recent theoretical calculations have shown that the optical and transport properties of QDs can be modified by external magnetic fields \cite{Singh_2017, PhysRevB.107.064420}, and the SOC's strengths can be tuned by gate electric fields and in-plane magnetic fields \cite{NittaPhysRevLett.78.1335, PhysRevB.83.245324}.
In particular, the manipulation of single and two-spin electron qubits can benefit from the generation and quantification of individual electron correlations \cite{ a16dbae8cf944942b1516c1a932241b4, bellCorrelations}. The potential transformation of semiconductor structures into bits of quantum information processing, depends on the ability to trap electrons in QDs and regulate their quantum states \cite{PhysRevA.57.120}. Studies in optical spectroscopy have shed light on related processes \cite{cavQdot1, PhysRevResearch.3.043161} leading to the presence of stable topological properties such as spin vortices \cite{ LuoNature, Chakraborty}. However, a microscopic knowledge of the spatially inhomogeneous QD electronic spin state, which is essential for regulating their mesoscopic behavior, is still lacking.

Experiments have demonstrated the successful coupling of charge and spin degrees of freedom in semiconductor QDs with visible and microwave photon cavities \cite{cavQdot1, PhysRevResearch.3.043161}. In particular, the polarization (linear or circular) of confined light can be controlled in an efficient way by means of semiconductor-superconductor cavities \cite{ PhysRevResearch.3.043161, RevModPhys.93.025005,  Owens2022}, demonstrating anticorrelation electron effects. The main focus is to understand electron spin textures in semiconductor nanostructures \cite{ PRXQuantum.4.010329}, as well as achieving high-sensitivity spin sensing using tip-based technologies such as scanning tunneling microscopy (STM) \cite{qdotSTM, rodary:hal-02337790}. In particular, double-tip STM \cite{twoProbeSTM} and spin-polarized STM (SP-STM) are powerful techniques that can explore the correlation between electrical and magnetic properties with atomic resolution, and provide measurements to probe non-local spatial spin electron correlations \cite{Eltschka2014}. As a result, the non-local spin-photon optical properties in an hybrid QD-cavity system effectively open the way to studying quantum gates for quantum information.


In this paper, we analyze the interplay between the electron SOC, the magnetic field, and the cavity coupling in a QD system comprising two interacting electrons. Though obviously simplified in terms of real material complications, our purpose is precisely to focus in on the interplay of these terms -- and just these terms -- in the system's Hamiltonian and hence its quantum properties. In the presence of dominant Dresselhaus SOC and mixed Dresselhaus-Rashba SOC, our main interest is on the quantum correlations between matter and light states. We show that SOC causes the development of new polariton modes as a result of the decoupling between the cavity and the two-electron center of mass (CM) mode. By adjusting the magnetic field and the cavity coupling, we tune the spin states into resonance, and demonstrate the entanglement of the relative modes with the cavity. We also demonstrate how tuning the cavity coupling controls the distribution of the Bell spatial states and the spin correlation for the two electrons. 

We show that: (1) highly spin-topological properties can be tailored by the microwave cavity field; (2) the violation of the Kohn theorem (KT) \cite{KTheorem} is shown to be more pronounced in the USC regime and can be adjusted by the microwave cavity field; and (3) besides their fundamental interest, our results demonstrate that measurements of the ground state (GS) transitions enable tuning of correlated spins and the generation of spatially inhomogeneous Bell states. 

The paper is organized as follows. Firstly, in Section \ref{sec:teoQEDSOC} the basic theoretical background is provided for incorporating hybrid SOC-cavity effects in an interacting two-electron QD, focusing on the central quantities of interest: the charge density and the spin correlations. Subsequently, in Section \ref{sec:SpinField} we present the results for two cases: (A) Dresselhaus SOC alone and (B) mixed Dresselhaus-Rashba SOC. Finally, in Section \ref{sec:Bell} we discuss the numerical resolved spin Bell state distribution in a QD. Our main conclusion are summarized in Section \ref{sec:conclusion} and the Appendices contain particular details of our calculations.
\vspace{-0.3cm}
\section{Theoretical framework: QED-SOC Hamiltonian}\label{sec:teoQEDSOC}
\begin{figure}[h]
    \centering
    \includegraphics[width=0.45\textwidth]{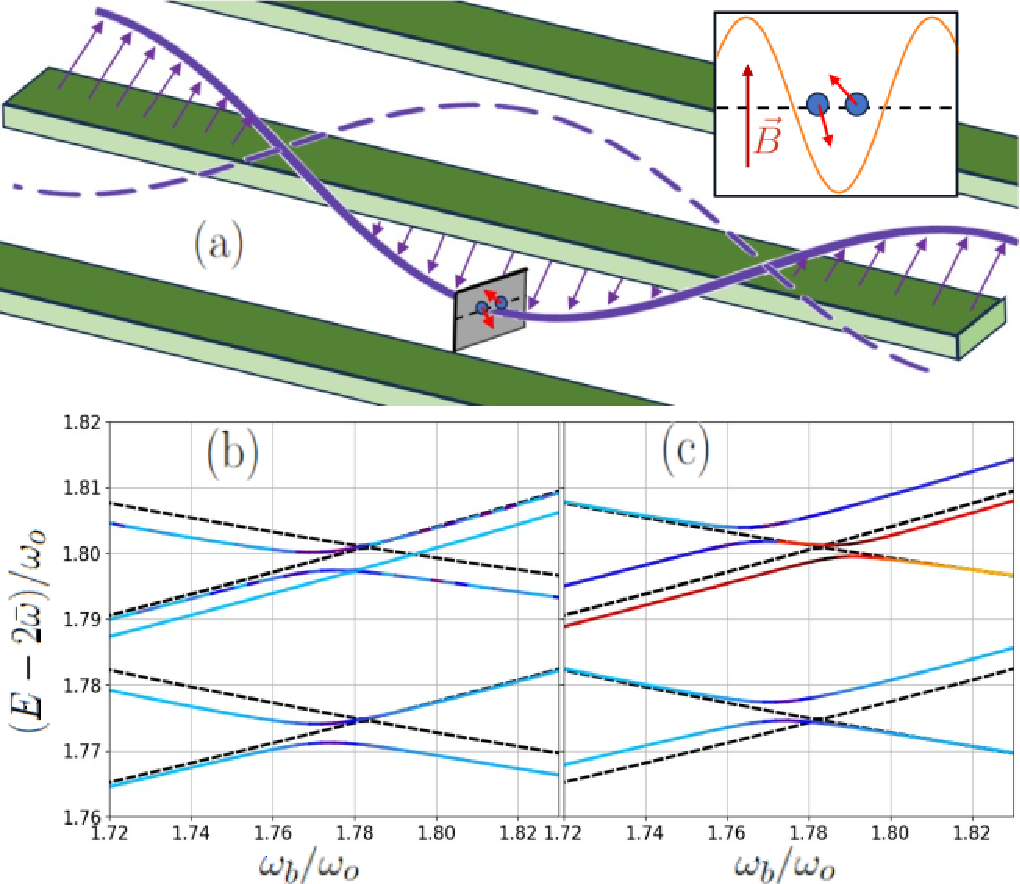}
    \caption{(a) Schematic of a quantum dot (QD) with two interacting electrons  (inset) inside a superconducting microwave cavity with circularly polarized light (CPL) \cite{RevModPhys.93.025005,  Owens2022}. Perpendicular external magnetic and electric fields  modify the strength of the spin-orbit-coupling (SOC). Energy spectrum as a function of the cyclotron frequency at $\lambda_D=0.04$ (b) without cavity coupling; (c) with cavity coupling $g=0.07$. Dashed curves represent the non-interacting case ($\lambda_D=\lambda_R=0$). The relative mode SOC splitting (or the photonic SOC splitting) is indicated by more blue colors (or more red colors). A curve with a more purple color (or more maroon color) means the state is more mixed.}
    \label{fig:cavInteracciones}
\end{figure}

The hybrid QD-cavity system is schematically shown in Fig. \ref{fig:cavInteracciones}a. Two electrons are confined in a parabolic two-dimensional QD embedded in a single-mode cavity, in the presence of a  perpendicular magnetic field $B\hat{z}$. The Hamiltonian of the total system (with $c=\hbar=1$):
\begin{subequations}
    \begin{align}
\mathcal{H}=&\sum_{j=1}^2\{ \mathcal{H}_{o}^{(j)}+\mathcal{H}_{soc}^{(j)}\}+\omega_c a^\dagger a+V_{e-e}(|\vec{r}_1-\vec{r}_2|),\label{eq_H_inicial}\\
\mathcal{H}_{o}^{(j)}=&\frac{|\vec{\Pi}_j|^2}{2m^*}+\frac{m^*}{2}\omega_o^2r_j^2+\frac{\Delta_Z}{2} \sigma_z^{(j)},\\
     \mathcal{H}_{soc}^{(j)}=&{\lambda_D}(\sigma_{y}^{(j)}\Pi_{j,y}-\sigma_{x}^{(j)}\Pi_{j,x})+{\lambda_R}(\sigma_{x}^{(j)}\Pi_{j,y}-\sigma_{y}^{(j)}\Pi_{j,x}), \label{eq_hsoc_j}
        \end{align}
    \end{subequations}
where $\sigma_k^{(j)}$ with $k=\{x,y,z\}$ refers to a Pauli matrix for particle $j$. 

The harmonic QD is characterised by energy $\omega_o$ and length $\ell_o = [1/(m^*\omega_o)]^{1/2}$. The electron-electron interaction, usually taken as the Coulomb repulsion, is here incorporated through an effective potential $V_{e-e}(|\vec{r}|)= \frac{\alpha}{r^2}$ (where $\vec{r}=\vec{r}_1-\vec{r}_2$ is the two-electron relative coordinate and $\alpha$ is an adjustable parameter to reproduce the Coulomb results), whose higher-order power mimics the presence of mirror charges in the semiconductor QD. This effective dipole-like interacting two-electron model has attracted significant interest in the literature since, despite its simplicity, it yields nontrivial results which are comparable with the bare Coulomb interaction results. Without including SOC and cavity effects, an analytical solution in the presence of an external magnetic field is obtained \cite{Quiroga1993SpatialCO}, which depends on Landau levels $n_{cm}$ ($n_r$) and angular momentum quantum number $m_{cm}$ ($m_r$) of CM (relative) space and yields analytic matrix elements, improving the efficiency of numerical results and allowing the use of numerical exact diagonalization. Moreover, this model has been extended to explore new physical results in multi-electron QDs \cite{ PhysRevLett.67.1157, Johnson1995AnalyticRF}, Quantum Hall systems \cite{Johnson_1997, PhysRevB.46.4681} as well as nuclear-electron quantum logic gates \cite{PhysRevB.62.R2267}.

The external magnetic and cavity fields are included through the vector potentials which determine the kinetic momentum $\vec{\Pi}_{j}=\vec{p}_{j}+e\vec{A}^{(c)}_{j}+e\vec{A}^{(q)}$. In what follows, we choose the symmetric gauge vector potential $\vec{A}^{(c)}_{j}=\frac{B}{2}(-y_j,x_j,0)$. The single-mode cavity radiation which couples to matter through the electric dipolar approximation, is represented by $\vec{A}^{(q)}=A(a^\dagger\vec{\epsilon}^*+a\vec{\epsilon})$ and is independent from the particle label $j$, and  $a(a^\dagger)$ is the annihilation (creation) operator of a single-mode photon with energy $\omega_c$. We consider a right circularly polarized light (CPL) field as specified by $\vec{\epsilon}=\frac{1}{\sqrt{2}}(1, i,0)$. In addition, the magnetic field causes Zeeman splitting which is given by $\Delta_Z=g_L \mu_B B$, where $g_L$ is the Landé factor and $S_z=\frac{1}{2}(\sigma_z^{(1)}+\sigma_z^{(2)})$.

Eq. \ref{eq_hsoc_j} describes the interaction between kinetic momentum and spin for both electrons, encompassing the primary linear Dresselhaus and Rashba interactions with parameters $\lambda_D$ and $\lambda_R$ respectively. As outlined in Appendix \ref{ap:bosonic}, a bosonic representation of these interactions is provided by 
\begin{subequations}
    \begin{align}
    \begin{split}
      \mathcal{H}_{soc}^{rel}=&-i\frac{\lambda_D}{2}\left(c_-\alpha_L+c_+\alpha_R^\dagger\right)\Sigma_{+}\\&\quad \quad \quad +\frac{\lambda_R}{2}\left( c_-\alpha_L^{\dagger}+c_+\alpha_R \right)\Sigma_{+}+ \text{h.c.},\label{eq:soc_rel} 
    \end{split}\\
\begin{split}
    \mathcal{H}_{soc}^{cm}=&-i\frac{\lambda_D}{2}\left(c_-a_L+c_+a_R^\dagger\right)S_{+}\\
      &\quad \quad \quad+ \frac{\lambda_R}{2}\left( c_- a_L^{\dagger}+c_+ a_R \right)S_{+}+\text{h.c.},\label{eq:soc_cm}
\end{split}\\ 
        \mathcal{H}_{soc}^{ph}&=-\eta(g)(\lambda_D a^\dagger S_+-i\lambda_R a S_+)+\text{h.c.},\label{eq:soc_ph}
    \end{align}
    \label{eq:soc_total}
\end{subequations}
\noindent where $c_\pm=\sqrt{\frac{\bar{\omega}\omega_o}{2}}\left(\frac{\omega_b}{2\bar{\omega}}\pm 1\right)$, taking the renormalized frequency $\bar{\omega}=\frac{1}{2}\sqrt{\omega_b^2+4\omega_o^2}$ and cyclotron frequency $\omega_b=\frac{eB}{m^*}$, and  h.c. is the hermitian conjugate. The bosonic operators $\alpha_R$ and $\alpha_L$ correspond to the annihilation operators for right and left relative oscillators, respectively, while $a_R$ and $a_L$ do the same for the correspondent CM oscillators. Specifically, $\alpha_R$ and $\alpha_L^\dagger$ ($a_R$ and $a_L^\dagger$) annihilate relative (CM) angular momentum states, as described in Appendix \ref{ap:bosonic}, thereby elucidating their coupling with spin. The cavity coupling constant is $\eta(g)=\frac{1}{\sqrt{2}}m^*g\ell_o$, with $g=eA\sqrt{\frac{\omega_o}{m^*}}$. Ladder spin operators $S_\pm=\sigma_{\pm_1}+\sigma_{\pm_2}$ and $\Sigma_\pm=\sigma_{\pm_1}-\sigma_{\pm_2}$ define transitions in which the eigenvalue $s$ of total spin projection $S_z$ is modified by one. In particular, $S_\pm$ affect exclusively triplet spin states transitions. The interplay between CM and light states, along with spin, leads to a decrease in energy for the GS with a higher contribution from triplet states. Conversely, $\Sigma_\pm$ is the responsible for transitions between singlet and triplet states such as $\ket{\uparrow\uparrow}$ or $\ket{\downarrow\downarrow}$ with simultaneous changes in the relative states.

Only the CM degrees of freedom are connected to the cavity due to dipolar interactions (see Appendix \ref{ap:bosonic}). However, nonzero SOC violates this idea since it introduces a link between relative modes and light. Resonance between photon modes and the spin Zeeman splitting is required ($\omega_c=|\Delta_z|$), which makes this violation stronger. In the presence of electron-electron interactions, the relative energy levels of the non-interacting case ($g=\lambda_D=\lambda_R=0$) exhibit crossings due to the conservation of total angular momentum $L_z$. Small energy differences between relative states thus resonate with spin and radiation (indirectly). Additionally, cavity-mediated SOC introduces interactions between the resonant spaces and motivates our investigation of the relatively unexplored optical properties, where the allowable transitions are determined by the SOC type and the type of CPL.

The non-interacting crossings, depicted by the dashed curves in Figs. \ref{fig:cavInteracciones}b and \ref{fig:cavInteracciones}c, experience splittings due to cavity-mediated SOC among states sharing the same conserved quantity (see Appendix \ref{ap:symmetries}). Specifically, when considering only the Dresselhaus SOC, the first terms in Eq. \ref{eq:soc_total} result in the conserved quantity $J_D=N+L_z-S_z$, where $N$ represents the photon number operator.
Eq. \ref{eq:soc_rel} gives rise to the two splittings (shown by more blue colors in Fig. \ref{fig:cavInteracciones}b). The Fock light states $\ket{0}$ and $\ket{1}$ have weights in the ground and excited states that are involved in the anticrossings, respectively. Moreover, interactions with CM states produce asymmetries in the splittings with respect to the non-interacting crossings (dashed curves). When states intersect along the blue curves at $\omega_b=1.78$ in Fig. \ref{fig:cavInteracciones}b, the cavity (through Eq. \ref{eq:soc_ph}) introduces an anticrossing depicted in more red colors and shifts the existing splittings.
Consequently, relative-spin-photon polaritons with $J_D=-1$ result from Dresselhaus SOC anticrossings in resonance. Larger state transitions are allowed by considering both SOCs, leading to splitting processes between states with different $J_D$ values.

In order to assess spin textures, we resort to the calculation of the spatial variation of total spin projections at two diametrically opposite positions around the QD origin. This means that the CM position is fixed at $\vec{R}=\frac{1}{2}(\vec{r}_1+\vec{r}_2)=0$, while $\vec{r}$ is varied. This way of detection allows us to observe changes in relative coordinates due to the cavity. Specifically, the relative density and spin field of the ground state $\ket{GS}$ are defined respectively as:
\begin{subequations}
    \begin{align}
        \rho(\vec{r}):=&\Tr_{ph, s}(|GS\rangle\langle GS|)|_{\vec{R}=0},\\
        \mathcal{S}_{k}(\vec{r}):=&\Tr_{ph, s}\left(|GS\rangle\langle GS|S_k\right)|_{\vec{R}=0},
    \end{align}
\end{subequations}
where the traces run over the spin and photon spaces and total spin in the $k$ direction: $S_k=\frac{1}{2}(\sigma^{(1)}_k+\sigma^{(2)}_k)$. The photon density $\rho_{ph}=\sum_{i,j} C_{ij} |n_i\rangle \langle n_j|$ is obtained by tracing over matter states. Bell state densities consider the spin states: singlet $\ket{\phi_-}=\ket{A}=\frac{1}{\sqrt{2}}(\ket{\uparrow\downarrow}-\ket{\downarrow\uparrow})$, the symmetric state of the triplet $\ket{\phi_+}=\ket{S}=\frac{1}{\sqrt{2}}(\ket{\uparrow\downarrow}+\ket{\downarrow\uparrow})$, and the linear combinations of triplet states $\ket{\psi_\pm}=\frac{1}{\sqrt{2}}(\ket{\uparrow\uparrow}\pm\ket{\downarrow\downarrow})$.

\begin{figure*}
    \centering
    \includegraphics[width=0.98\textwidth]{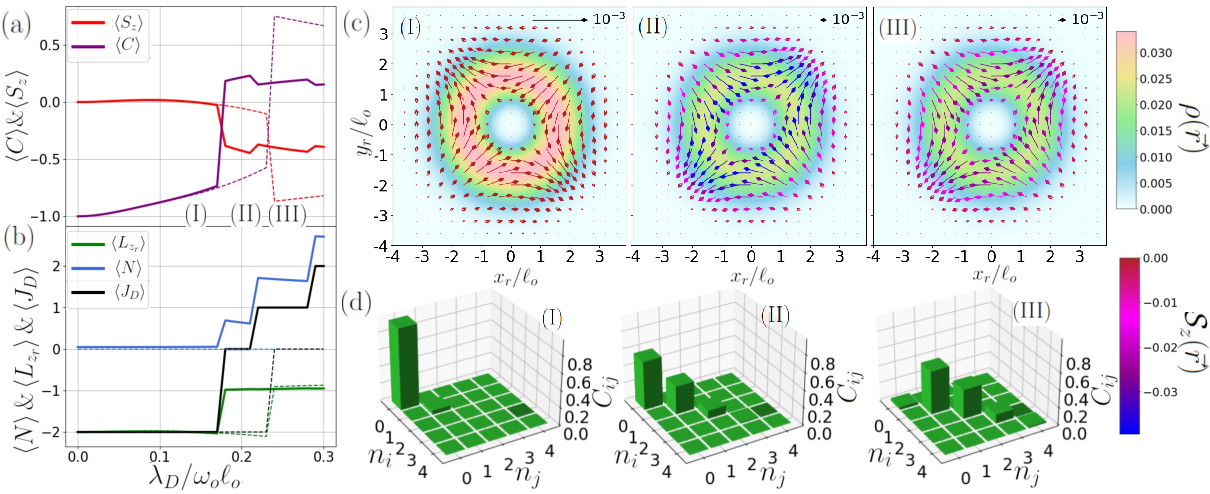}
    \caption{(a) $\langle C\rangle$, $\langle S_z\rangle$ and 
 (b) $\langle N\rangle$, $\expval{L_{z_r}}$ and $\expval{J_D}$ as function of $\lambda_D$ at $\alpha=3$ and fixed $g=0.28 $ and $g=0.00$ (dashed). (c) Relative density $\rho(\vec{r})$ with pastel colors, and spin field illustrated by arrows for its in plane components $(\mathcal{S}_{x}(\vec{r}), \mathcal{S}_{y}(\vec{r}))$ and by colors for $\mathcal{S}_{z}(\vec{r})$ at (I) $\lambda_D=0.15$, (II) $\lambda_D=0.21$ and (III) $\lambda_D=0.25$. (d) The associated photon density matrix $\rho_{ph}$ according to those coupling values, where the bar height represents $C_{i,j}$.}
    \label{fig:densityCSP_Dressel}
\end{figure*}

\section{Spin field controlled by SOC-cavity}\label{sec:SpinField} 

As a concrete example, we consider a two-electron GaAs QD, with relatively small Zeeman interaction but strong SOC,
immersed in a microwave cavity. The values of the QD parameters are as follows: $m^* = 0.067m_e$ is the electron effective mass, $g_L = -0.44$ is the Landé g-factor \cite{PhysRevB.38.1806} and $\ell_o = 50$ nm ($\omega_o=0.45$meV)  
is the confinement length. Cyclotron frequency $\omega_b = 1.9$ is chosen and SOC strength is changed from zero to $\lambda = 0.3$ \cite{PhysRevB.38.1806}. Cavity coupling is tuned up to $g = 0.3$, which corresponds to the USC regime with $g/\Omega \sim 1$. $\Omega=\omega_c+2\frac{g^2}{\omega_o}$ is the effective radiation frequency associated with the diamagnetic term $|\vec{A}^{(q)}|^2$. SOC effects from Dresselhaus interaction ($\lambda_D\neq 0$), which is characterized by a conserved quantity, and the Dresselhaus-Rashba combination ($\lambda_D=\lambda_R \neq 0$) are discussed separately using a numerically exact diagonalization. 

Energy level convergence was ensured through a two-step process. Firstly, a basis of uncoupled oscillators was formulated for the Hamiltonian, taking into account the dipolar interaction with the CM coordinates through a Bogoliubov transformation, which is explained in Appendix \ref{ap:CMP_polariton}, and the maximum quantum numbers were specified. Secondly, the Hamiltonian matrix was reorganized into blocks based on conserved quantities, which provided us with both computational and analytical advantages. In fact, the ground state's optoelectronic properties exhibit distinct trends due to the presence of symmetries.
\vspace{-0.5 cm}

\subsection{Dresselhaus SOC}\label{ssec:DresDominant}

The presence of a conserved quantity and the coupling variation, result in transitions in the ground state characterized by integer values of $J_D$ (see Fig. \ref{fig:densityCSP_Dressel}b). These transitions are associated with simultaneous changes in $\langle N \rangle$, $\langle S_z \rangle$ and $\langle L_{z_r} \rangle$, as shown in Figs. \ref{fig:densityCSP_Dressel}a and \ref{fig:densityCSP_Dressel}b. The control of spin transitions is achieved through the strength of Dresselhaus SOC ($\lambda_D$) and cavity coupling ($g$), and their detection is linked to changes in the light states. In Fig. \ref{fig:densityCSP_Dressel}a, significant spin transitions are observed at low interactions and are detected through discontinuities in $\langle S_z \rangle$ and spin correlations $\langle C \rangle$, in which $C=\sum_{s_1,s_2}s_1s_2 \ket{\varsigma_1,\varsigma_2}\bra{\varsigma_1,\varsigma_2}$ for $\varsigma=\uparrow(\downarrow)$ and $s_1=1(-1)$. Subsequent transitions become less prominent in spin but significant in $\langle N \rangle$, resulting in a staircase-like behavior of $\langle C \rangle$ caused by the presence of $J_D$. These correlations move toward uncorrelated spins with particular average values of $\langle S_z\rangle\sim -0.40$ and $\langle C\rangle\sim 0.18$, indicating an enhancement of spin entanglement with matter and radiation.

\begin{figure*}
    \centering
    \includegraphics[width=\textwidth]{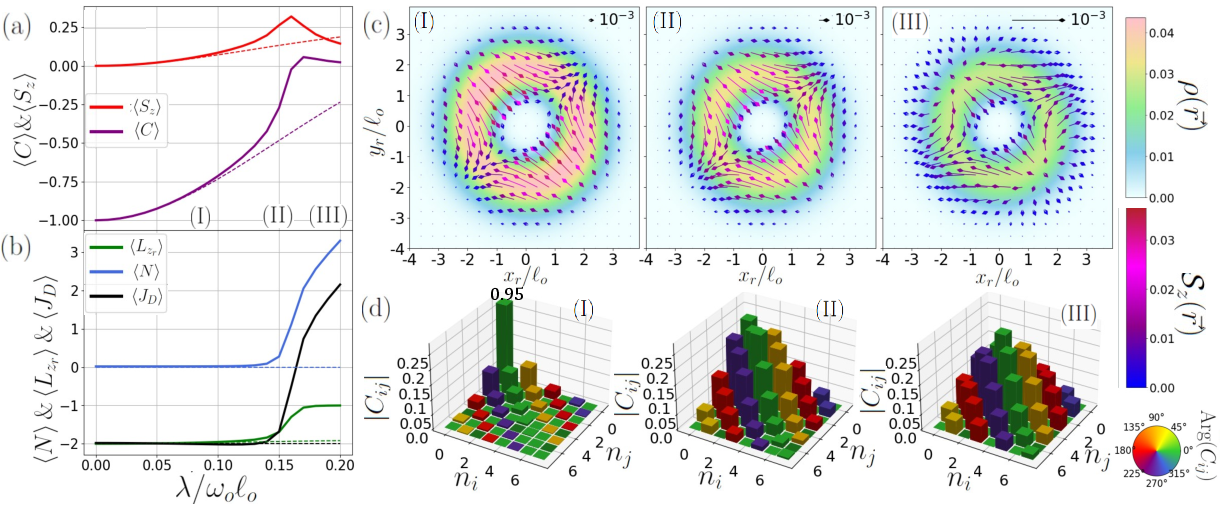}
    \caption{(a) $\langle C\rangle$, $\langle S_z\rangle$ and 
 (b) $\langle N\rangle$, $\expval{L_{z_r}}$ and $\expval{J_D}$ as function of $\lambda_D=\lambda_R=\lambda$ at $\alpha=3$ and fixed $g=0.18 $ and $g=0.00$ (dashed). (b) Relative density $\rho(\vec{r})$ with pastel colors, and spin field illustrated by arrows for its in plane components $(\mathcal{S}_{x}(\vec{r}), \mathcal{S}_{y}(\vec{r}))$ and by colors for $\mathcal{S}_{z}(\vec{r})$ at (I) $\lambda=0.08$, (II) $\lambda=0.15$ and (III) $\lambda=0.19$. (c) The associated photon density matrix $\rho_{ph}$ according to those coupling values, where the bar height represents $|C_{i,j}|$ and bar colors refers to $\text{Arg}(C_{ij})$.}
    \label{fig:densityCSP_DresRas}
\end{figure*}

In Fig. \ref{fig:densityCSP_Dressel}c (I) to (II), a $90^\circ$ rotation in the spin field is observed, accompanied by a change in the $z$-component from zero to negative values, represented by a change in the color of the arrows from red to blue. This transition occurs simultaneously with an increase in the weight of the projector $\ket{1}\bra{1}$ in the photon occupation matrix weights, as depicted in Fig. \ref{fig:densityCSP_Dressel}d. The next transition in the spin field between (II) and (III) happens only in its $z$-component, due to a small discontinuous variation in $\langle S_z\rangle$ and $\langle C\rangle$. These discontinuities are detected by $\langle N\rangle$, where an increase in the weight of the projector $\ket{2}\bra{2}$ is observed when comparing Figs. \ref{fig:densityCSP_Dressel}d (II) and (III). Therefore, transitions are observed in charge and spin densities, which retain Dresselhaus-like spin fields \cite{LuoNature} and demonstrate a Gaussian electronic density centered on a radius with a reduction in the probability of finding the QD electrons at $\vec{R}=0$, which is due to the couplings with CM states.

The transition from the singlet state $\ket{m_r=-2, \phi_-}$ (anti-correlated) to uncorrelated spins is made possible by the cavity-SOC coupling. The interactions described by Eqs. \ref{eq:soc_rel} and \ref{eq:soc_ph} enhance the probabilistic weight of triplet states as the role of transitions involving light states increases. In fact, a polariton state characterized by a higher contribution from states $\ket{m_r=-1, \phi_+}$ and $\ket{m_r=-1, \downarrow\downarrow}$ is formed, accompanied by an increase in photon occupation and $\expval{L_{z_r}}$ transition. When the coupling between CM modes and photon states increases, mediated by spin, this state becomes the new GS (see Appendix \ref{ap:secondPert}). 

In the absence of electron-electron interactions, the system exhibits a vacuum light state with no ground state crossings, because there is no distinction in terms of energy between CM and relative states. However, when the electron-electron interaction and a magnetic field are present, the energy spectrum is modified, which affects the required couplings for light-matter transitions. With the Coulomb potential turned on, in Fig \ref{fig:densityCSP_Dressel}c, a mixed light state with diagonal photon matrix emerges, due to the resonance condition. These discoveries provide tangible evidence for violating the Kohn Theorem, since variations in the relative parameters directly impact the light states, which in turn enables their detection through methods like state tomography or Wigner function measurements. 

\vspace{-0.3 cm}

\subsection{Rashba-Dresselhaus SOC}\label{ssec:DresRasDominant}

The coexistence of Rashba and Dresselhaus SOCs gives rise to new counter-rotating terms connecting light and spin states, as indicated by the second term in Eq. \ref{eq:soc_ph}. Similarly, all the interactions between the CM and relative coordinates of matter with spin must be evaluated, as described by Eqs. \ref{eq:soc_cm} and \ref{eq:soc_rel}. 
In the presence of strong couplings, the splitting effects are amplified, resulting in increased entanglement between matter and radiation spin states. This entanglement is evident in the higher photon number $\langle N \rangle$ observed at lower couplings ($0.15 <\lambda <0.20$ and $g=0.18$) in Fig. \ref{fig:densityCSP_DresRas}b compared to Fig. \ref{fig:densityCSP_Dressel}b at $g=0.28$. Additionally, the photon densities in Fig. \ref{fig:densityCSP_DresRas}d exhibit off-diagonal terms and higher Fock states, indicating a greater possibility of transitions with light states and providing empirical evidence for spin-radiation virtual transitions.

\begin{figure*}
    \centering
    \includegraphics[width=\textwidth]{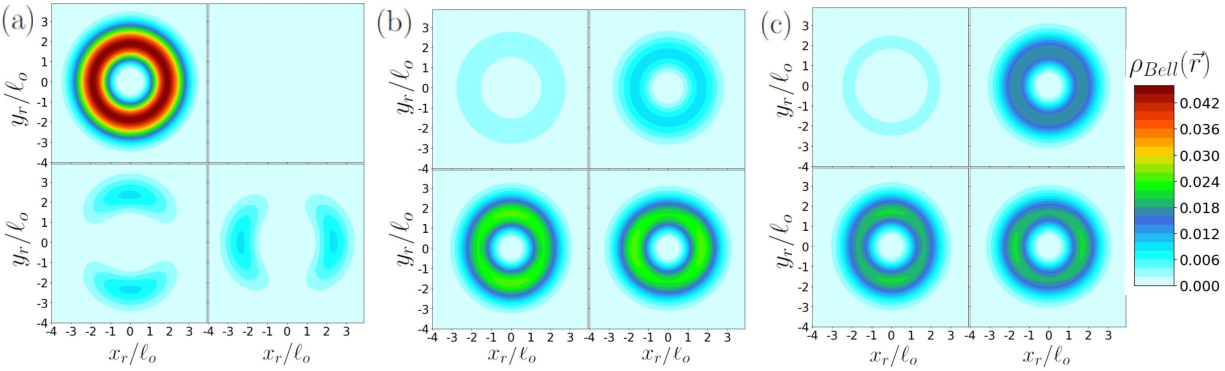}
    \caption{Spin Bell states' distribution $\rho_{Bell}(\vec{r})=\mathcal{N}\Tr_{ph,s}{(\ket{GS}\bra{GS}\ket{B}\bra{B})}|_{\vec{R}=0}$, normalized at $\vec{R}=0$, with respect to the relative coordinates at $\lambda_D=0.22$ and $\alpha=3$, and (a) $g=0.08$, (b) $g=0.16 $ and (c) $g=0.30$. Bell states ($\ket{B}$) shown are $\ket{\phi_-}$ (upper left plot), $\ket{\phi_+}$ (upper right plot), $\ket{\psi_-}$ (lower left plot) and $\ket{\psi_+}$ (lower right plot).}
    \label{fig:bellDistri_Dres}
\end{figure*}

\begin{figure*}
    \centering
    \includegraphics[width=\textwidth]{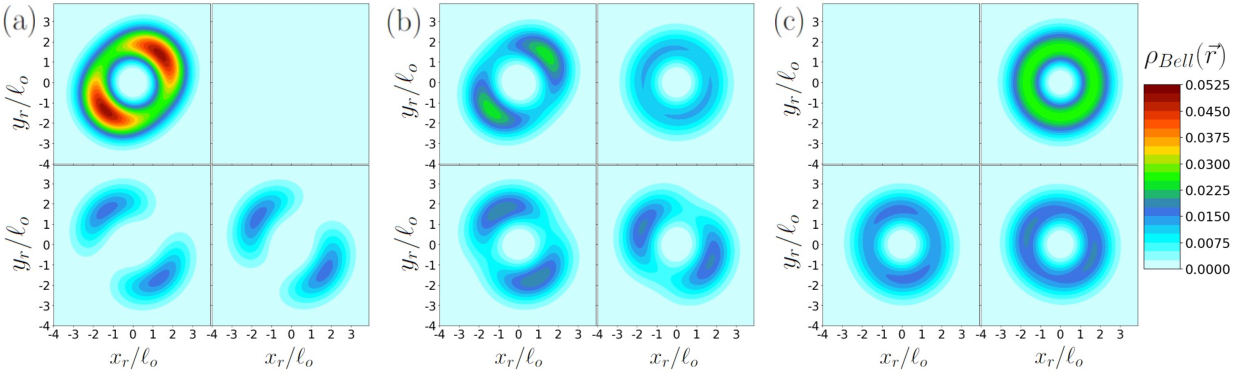}
    \caption{Spin Bell states' distribution $\rho_{Bell}(\vec{r})$ with respect to the relative coordinates at $\lambda=\lambda_D=\lambda_R=0.18$ and $\alpha=3$, and (a) $g=0.12$, (b) $g=0.16$ and (c) $g=0.18$. The same distribution of Bell states is shown as in Fig. \ref{fig:bellDistri_Dres}.}
    \label{fig:bellDistri_DresRas}
\end{figure*}

Figs. \ref{fig:densityCSP_DresRas}a and \ref{fig:densityCSP_DresRas}b exhibit smooth variations in the expected values as a consequence of the absence of a conserved quantity \cite{LuoNature, Chakraborty}. A detailed examination of the changes in cavity coupling (see Appendix \ref{ap:cavDep}) further reveals a trend towards uncorrelated spins and $\langle S_z \rangle \sim 0$, accompanied by an increase in $\langle N \rangle$. Moving from Fig. \ref{fig:densityCSP_DresRas}c (I) to (II) results in a reorientation of spin densities, with their vortex points shifting from $\pm(2.4,2.4)$ to $\pm(2.7, 2.7)$, and notable alterations in the associated photon densities, characterized by variations in the matrix weights. Subsequently, transitioning from Fig. \ref{fig:densityCSP_DresRas}c (II) to (III) amplifies the divergent tendencies in the spin texture along the $y=0.5x$ axis, accompanied by the emergence of divergent vortical points at $\pm(1.3,-1.3)$. These modifications also augment the contribution of higher Fock light states in Fig. \ref{fig:densityCSP_DresRas}d.

Under the same couplings, the electron-electron interaction once again leads to different light states as compared to the vacuum state, which is observed when the radiation is off-resonance at $\alpha=0$. Due to the smooth shift from $\langle L_{z_r}\rangle\sim-2$ to -1 in Fig. \ref{fig:densityCSP_DresRas}b, the charge density also exhibits a reduction in the most probable relative radius of locating the electrons, from $r=1.6$ to $r=1.4$. This transition is caused by SOC in the present case, but the cavity is also able to induce it as demonstrated in Appendix \ref{ap:CoulombHole}. In fact, as a result of the Kohn Theorem violation, light can affect the repulsive effects that depend on relative coordinates.

\vspace{-0.2 cm}
\section{Spin Bell state textures}\label{sec:Bell}

The transitions in the QD's total spin field described in the previous sections provide evidence of the localization of spin Bell states. The possibility of observing spin-entangled states is controlled by tuning SOC strengths, due to resonance between spin and relative states. However, the presence of the cavity is crucial for shaping these spin states since it facilitates new transitions through jumps between Fock states. These transitions exhibit a staircase-like pattern of correlations (see Appendix \ref{ap:cavDep}) which depends on $g$ and leads to modifications in the densities of the Bell states.

The cavity successfully maintains the trend of $\langle S_z \rangle < 0$ due to the Dresselhaus SOC, enabling transitions at lower values of $\lambda_D$ (see Fig. \ref{fig:densityCSP_Dressel}a). Fig. \ref{fig:bellDistri_Dres}a illustrates how the singlet state begins to transition into a state with a higher weight of $\ket{\downarrow\downarrow}$. Then, in Fig. \ref{fig:bellDistri_Dres}b, the densities of the $\ket{\psi_-}$ ($\ket{\phi_+}$) exhibit an asymmetric emergence, with radial localization at $r=2$ and maximum values at angles of $0^\circ$ ($90^\circ$) caused by the mixture with relative modes. The cavity-mediated SOC also introduces contributions from other triplet spin states, resulting in uncorrelated spins and an increased probability density of finding the state $\ket{\phi_+}$ in Fig. \ref{fig:bellDistri_Dres}c. While it possesses a clear azimuthal symmetry, local spin transformations \cite{bellCorrelations} could enhance its strength and facilitate its detection.

Due to the lack of a conserved quantity in the presence of both SOCs, the densities of the Bell states show azimuthal asymmetry while spin correlations are smoothed and tend to zero as the couplings increase.
Transitions to the $\ket{\downarrow\downarrow}$ and $\ket{\uparrow\uparrow}$ (caused by Rashba SOC) are facilitated by the cavity at lower $\lambda$ despite the SOCs competition. As seen in Fig. \ref{fig:bellDistri_DresRas}b, the singlet state is concentrated at coordinates $\pm (1.75,- 1.75)$, and with increased $g$-coupling, the probability to  get $\ket{\psi_\pm}$ states improves considerably.
In Fig. \ref{fig:bellDistri_DresRas}b, the non-homogeneous localization of these states is concentrated in distinct peaks $\pm(1, -1.6)$ for $\ket{\psi_-}$ and $\pm(1.6, -1)$ for $\ket{\psi_+}$. The different localization within the QD is attributed to the exclusive combination of each $\ket{\psi_\pm}$ states with light and relative states, demonstrating spatial spin correlations in microcavities. 
In Fig. \ref{fig:densityCSP_DresRas}c, the probability of finding the $\ket{\phi_+}$ state increases, exhibiting an azimuthally symmetric density concentrated at $r=1.5$. 
The presence of triplet states at large $g$ is explained by the direct coupling of light with these states in Eq. \ref{eq:soc_ph}, and it is consistent with the trend towards uncorrelated spins. Additionally, densities demonstrate a loss of asymmetry effect, as illustrated also in Appendix \ref{ap:CoulombHole}, revealing an increasing of confinement as a result of strong couplings with the cavity and mixture with a greater number of light states.

\vspace{-0.2 cm}
\section{Summary and Discussion}\label{sec:conclusion}

The interaction of a two-electron QD with cavity coupling, magnetic fields, and spin-orbit coupling has been investigated in order to understand how these factors contribute to the formation of polariton states and complex spin-spatial correlation inhomogeneities.
The cavity-mediated spin-orbit-coupling enables interactions between the center-of-mass and relative modes of the cavity, resulting in mixed light-matter states. These states provide measurable evidence of Kohn's Theorem violation, where the Coulomb hole is modified by the cavity and electron-electron interactions alter the light states. Furthermore, the control of spin correlations can be attributed to the mediation in the radiation-matter coupling. 

Dresselhaus spin-orbit-coupling, on one hand, gives rise to staircase-like spin correlations due to its connection with transitions involving the conserved quantity $J_D$. On the other hand, the inclusion of Rashba spin-orbit-coupling leads to the smoothing of correlations and a tendency towards uncorrelated spins, caused by the inclusion of virtual spin-radiation transitions and subsequent interactions with matter. The change in correlations is linked to the generation of Bell states with inhomogeneous spatial distributions. In this way, quantum dots in microwave cavities can potentially be used as an experimental platform for full solid-state Bell tests \cite{Bordoloi_2022}, where the detection of light states reflects the spin properties. Additionally, the flexible control and measurement of spin states contribute to the development of protocols for quantum information processing with few-particle systems, the generation of entangled systems, and applications in spintronics \cite{spintronics}.

\section*{Acknowledgments}
S.S.B-R, F.J.R. and L.Q. are thankful for financial support from Facultad de Ciencias-
UniAndes Projects No. INV2021-128-2292 and INV-2023-162-2833, and 
Universidad de los Andes High Performance Computing (HPC)
facility. N.F.J. is supported by U.S. Air Force Office of Scientific Research awards FA9550-20-1-0382 and FA9550-20-1-0383.

\appendix

\section{QED-SOC Hamiltonian: CM and relative coordinates}\label{ap:bosonic}

The introduction of CM and relative coordinates is motivated by the direct coupling of light in the cavity with the CM states and the analytical solution to the relative part in the presence of an external magnetic field and a potential $\alpha/r^2$ \cite{Quiroga1993SpatialCO}. In addition, the bosonic version of our Hamiltonian requires the definition of operators that describe the interaction between $B\hat{z}$ and two electrons in the QD, without electron-electron interaction, as:
\begin{subequations}
    \begin{align}
        a_h=&\frac{\beta_{cm}}{{2}}\left[(R_x+i\eta R_y)+\frac{i}{\beta^2_{cm}}(p_{cm_x}+i\eta p_{cm_y})\right],\\
        \alpha_h=&\frac{\beta_{r}}{{2}}\left[(r_x+i\eta r_y)+\frac{i}{\beta^2_{r}}(p_{r_x}+i\eta p_{r_y})\right],
    \end{align}
\end{subequations} 
where the right (left) oscillator corresponds to $h=R(L)$ and $\eta=-1(1)$. $r_j$ and $R_j$ are the relative and CM positions, and its respective conjugate momentum are $p_{r_j}$ and $p_{cm_j}$. In this case $\vec{r}=\vec{r_1}-\vec{r_2}$, $\vec{R}=\frac{1}{2}(\vec{r_1}+\vec{r_2})$, $\vec{p}_{cm}=\vec{p_1}+\vec{p_2}$, and $\vec{p}_{r}=\frac{1}{2}(\vec{p_1}-\vec{p_2}).$ In addition, we fix the variables $\beta_{cm}=2\beta_r=\sqrt{2m^*\bar{\omega}}$. These operators define the creation of particles whose momentum and position variables follow either a clockwise or counterclockwise rotation. In their creation operator form, $a^\dagger_R$ ($\alpha^\dagger_R$) defines counterclockwise rotations in the CM (relative) phase space, while $a^\dagger_L$ ($\alpha^\dagger_L$) defines clockwise rotations in their respective spaces. This will be consistent with the creation of states that increase their angular momentum by $\hbar$ for $a^\dagger_R$ or $\alpha^\dagger_R$ and decrease it by $\hbar$ for $a^\dagger_L$ or $\alpha^\dagger_L$, as will be described later.

In essence, our Hamiltonian is rearranged as follows:
\begin{equation}
\begin{split}
\mathcal{H}=&\mathcal{H}_o^{cm}+\mathcal{H}_o^{rel}+\mathcal{H}_f+\Delta_Z S_z+\mathcal{H}_{dip}\\
&+\mathcal{H}_{soc}^{cm}+\mathcal{H}_{soc}^{rel}+\mathcal{H}_{soc}^{ph}.  
\end{split}
\label{eq:hamiltonian}
\end{equation}
The unperturbed Hamiltonian is composed by the Zeeman interaction, along with the harmonic oscillator Hamiltonians described by
\begin{subequations}
    \begin{align}
    \mathcal{H}_o^{cm}&=\omega_Ra_R^\dagger a_R+\omega_La_L^\dagger a_L+\Bar{\omega},\label{eq:h0_cm}\\
    \mathcal{H}_{o}^{r}=&\omega_R\alpha_R^\dagger \alpha_R+\omega_L\alpha_L^\dagger \alpha_L+\Bar{\omega}+\frac{\alpha}{r^2},\\
    \mathcal{H}_{ph}=&\Omega\left(a^\dagger a+\frac{1}{2}\right), \label{eq:Hph_2e}
    \end{align}\label{eq:Hoscillator}
\end{subequations}
where $\omega_R=\bar{\omega}+\frac{1}{2}\omega_b$, $\omega_L=\bar{\omega}-\frac{1}{2}\omega_b$. $n_R, n_L$ ($\tilde{n}_R, \tilde{n}_L$) are the associated CM (relative) quantum numbers. In fact, Landau levels and angular momentum quantum numbers are given by $n_{cm}=n_R$ and $m_{cm}=n_R-n_L$ ($n_r=\tilde{n}_R$ and $m_r=\tilde{n}_R-\tilde{n}_L$). Thanks to the analytical solution in relative coordinates \cite{Quiroga1993SpatialCO} a renormalization in $m_r$ worked with the $\frac{\alpha}{r^2}$ potential. Due to the wave function's antisymmetry, the following spin states are determined: the singlet state ($\ket{A}=\frac{1}{\sqrt{2}}(\ket{\uparrow\downarrow}-\ket{\downarrow\uparrow})$), and the triplet states ($\ket{\uparrow\uparrow}, \ket{S}=\frac{1}{\sqrt{2}}(\ket{\uparrow\downarrow}+\ket{\downarrow\uparrow}), \ket{\downarrow\downarrow}$). These states must satisfy the condition that even (odd) values of $m_r$ only permit singlet (triplet) spin states. Eq. \ref{eq:Hph_2e} refers to the photonic part with the renormalized frequency $\Omega=2\frac{g^2}{\omega_o}+\omega_c$ due to the diamagnetic term $|\vec{A}^{(q)}|^2$.

Regarding interactions, SOC defines a coupling between spin and angular momentum, through which possible transitions between spin states and independent states of relative, CM motion, and light are defined. These interactions become evident when expressed as:
\begin{subequations}
    \begin{align}
    \begin{split}
     &\mathcal{H}_{soc}^{rel}=-i\frac{\lambda_D \beta_{r}}{2}\left[\left(\frac{\omega_b}{2 \bar{\omega}}-1\right)\alpha_L+\left(1+\frac{\omega_b}{2 \bar{\omega}}\right)\alpha_R^\dagger\right]\Sigma_{+} \\
      &+\frac{\lambda_R \beta_{r}}{2}\left[\left(\frac{\omega_b}{2 \bar{\omega}}-1\right)\alpha_L^{\dagger}+\left(1+\frac{\omega_b}{2 \bar{\omega}}\right)\alpha_R \right]\Sigma_{+}+\text{h.c.},    
    \end{split}    \\
    \begin{split}
    &\mathcal{H}_{soc}^{cm}=-i\frac{\lambda_D \beta_{c m}}{4}\left[\left(\frac{\omega_b}{2 \bar{\omega}}-1\right)a_L+\left(1+\frac{\omega_b}{2 \bar{\omega}}\right)a_R^{\dagger}\right]S_{+}\\
      &+\frac{\lambda_R \beta_{c m}}{4}\left[\left(\frac{\omega_b}{2 \bar{\omega}}-1\right)a_L^{\dagger}+\left(1+\frac{\omega_b}{2 \bar{\omega}}\right)a_R \right]S_{+}+\text{h.c.},     
    \end{split}\\
    &\mathcal{H}_{SOC}^{ph}=g\sqrt{\frac{m^*}{2\omega_o}}(-\lambda_D a^\dagger+i\lambda_R a )S_++\text{h.c.},
        \end{align}
    \end{subequations}
where in this case $\lambda_D$ and $\lambda_R$ have $\omega_o\ell_o$ units. The ladder operators $\alpha_R$ and $\alpha_L$ represent the annihilation of right and left relative oscillators, while $a_R$ and $a_L$ perform the same role for the respective CM oscillators. In particular, the increment in relative angular momentum $m_r$ is accomplish through operators $\alpha_R^\dagger$ and $\alpha_L$, because they cause $(\tilde{n}_R, \tilde{n}_L)\rightarrow(\tilde{n}_R+1, \tilde{n}_L)=(m_r+1, n_r+1)$ and $(\tilde{n}_R, \tilde{n}_L)\rightarrow(\tilde{n}_R,\tilde{n}_L-1)=(m_r+1, n_r)$ respectively. Equivalentely happens with $a_R^\dagger$ and $a_L$ with the increment in the CM angular momentum $m_{cm}.$ Additionally, the spin ladder operators are defined as $S_\pm=\sigma_{\pm_1}+\sigma_{\pm_2}$ and $\Sigma_\pm=\sigma_{\pm_1}-\sigma_{\pm_2}$, which modifies by one the total spin quantum number. These operators allow for transitions between spin states, such as $S_+=\sqrt{2}(\ket{\uparrow\uparrow}\bra{S}+\ket{S}\bra{\downarrow\downarrow})$ and $\Sigma_+=\sqrt{2}(-\ket{\uparrow\uparrow}\bra{A}+\ket{A}\bra{\downarrow\downarrow})$. Thus, we observe that transitions in triplet states are related to transitions in CM or photon states, while the ones that affect the singlet state are combined with relative states transitions.

Dresselhaus and Rashba SOC define the way in which matter and radiation couple to the spin. Relative and CM states exhibit a similar coupling in terms of the creation or annihilation of spin states, allowing us to define the permitted transitions and rearrangement of energy levels. In particular, with cavity-mediated SOC, the light states also induce transitions between spin states, which depend on the simultaneous presence of SOC strength and the coupling with the cavity field $g$. This coupling is characterized as rotating (counter-rotating) between various Fock states and the spin triplet levels when Dresselhaus (Rashba) SOC is considered. For this reason, working with Rashba SOC provides insights into the effects of virtual transitions in light.


On the other hand, the dipolar interaction for right CPL in the cavity takes the following form:
\begin{equation}
    \begin{split}
\mathcal{H}_{dip}&=ig\sqrt{\frac{\bar{\omega}}{\omega_o}} \left[\left(1+\frac{\omega_b}{2 \bar{\omega}}\right)a_R^{\dagger} a+\left(1-\frac{\omega_b}{2 \bar{\omega}}\right)a_L^{\dagger} a^{\dagger}\right]+\text{h.c.}\label{eq:hdip_2e}
    \end{split}
\end{equation}
In contrast, when working with left CPL, the possible transitions between matter and radiation are modified, as they involve the transformation $a\rightarrow a^\dagger$. Hence, photonic SOC can also alter the interaction between light states and spin. The combination of SOC and dipolar interaction is responsible for the complex entanglement between matter and radiation states. Furthermore, the enhancement of one or the other strongly depends on the resonance conditions that are established. In particular, our interest in interconnecting relative states and light states requires overcoming the effects of dipolar interaction. This is possible, as shown in the main text, through the simultaneous resonance among the relative, spin, and radiation states.

\vspace{-0.3 cm}

\subsection{Symmetries and conserved quantities}\label{ap:symmetries}

We only focus on right CPL, because the left CPL is equivalent and linear polarization can be written as combination of CPLs. In fact, whatever combination of Rashba and Dresselhaus SOC is considered a Hamiltonian with left CPL can be transformed with parity $\mathcal{P}(B)=-B$, a rotation in spin space $\sigma_+\rightarrow i\cos(\theta) \sigma_+ +\sin(\theta) \sigma_-$ with $\tan(\theta)=\frac{\lambda_D^2-\lambda_R^2}{2\lambda_D\lambda_R}$ and a $g_L$ renormalization into a Hamiltonian with right CPL and the same SOCs. Similarly, the equivalence between a pure Rashba and a pure Dresselhaus Hamiltonian has been demonstrated before through a spin rotation ($\sigma_x\leftrightarrow \sigma_y$ and $\sigma_z\leftrightarrow -\sigma_z$) and $\mathcal{P}(g_L)=-g_L$ \cite{LossSO, LuoNature}.

The cavity also allows for a conserved quantity when considering only one CPL and only one SOC. This conserved quantity is given in general by $J=N\pm (L_z \pm  S_z)$, which combines the number of photon operator $N$, the total angular momentum $L_z$, and $S_z.$ In parentheses, we present the conserved quantity for pure Rashba (Dresselhaus) SOC with +(-) sign, while the sign outside the parentheses depends on the circular polarization. If the radiation has right (left) CPL, the sign is +(-). These quantities indicate the appearance of mixed states between light, matter, and spin.

Working with the interaction $\alpha/r^2$ facilitates the interpretation of SOC-induced interactions. When considering a conserved quantity $J$, a matrix block is created by projecting onto states that share the same $J$. Considering $\alpha/r^2$ interaction, the off-diagonal interaction terms arise solely from the SOC, whereas with the conventional Coulomb potential, interactions between relative Landau levels are typically present. In fact, both the physical interpretation of allowed transitions and the convergence of energy levels are favored when employing the inverse square potential. This latter effect is attributed to the fact that the convergence of Landau levels occurs at lower values when it depends solely on the SOC, as the absence of SOC would already guarantee diagonalization through its analytical solution \cite{Quiroga1993SpatialCO}, which is not the case with the Coulomb potential.

\vspace{-0.3 cm}

\subsection{Center of mass-photon polaritons: Bogoliubov transformation} \label{ap:CMP_polariton}

Another aspect to highlight regarding our Hamiltonian is that in the absence of SOC, the interaction between light and matter is limited to how radiation couples with the CM. In fact, this reduced Hamiltonian, given by $\mathcal{H}=\mathcal{H}_o^{cm}+\mathcal{H}_{ph}+\mathcal{H}_{dip}$ for right circularly polarized light, takes the following form:
\begin{equation}
 \begin{split}
    \mathcal{H}&=\omega_Ra_R^\dagger a_R+\omega_L a_L^\dagger a_L+ \Omega a^\dagger a +\omega_+(a_R^\dagger a+a_Ra^\dagger)\\
    &\quad\quad+ i\omega_-(a_L^\dagger a^\dagger-a_L a)+\Bar{\omega}+\frac{1}{2}(\Omega-\omega_c),\\
\end{split}   
\end{equation}
where we define $\omega_\pm=g\sqrt{\frac{\Bar{\omega}}{\omega_c}}\left(1\pm\frac{\omega_b}{2\Bar{\omega}}\right)$. The quadratic form in bosonic operators allows for a Bogoliubov transformation that decouples the oscillators. Therefore, a unitary transformation on bosonic operators is defined, where $\vec{a}=\mathcal{U}\vec{b}$, with $\vec{a}^T=(a_R, a_L, a, a_R^\dagger, a_L^\dagger, a^\dagger)$ and $\vec{b}^T=(b_1, b_2, b_3, b_1^\dagger, b_2^\dagger, b_3^\dagger)$. Consequently,
\begin{equation}
\mathcal{H}=\sum_{j=1}^3\omega_j b_j^\dagger b_j -\frac{\omega_c}{2}
\end{equation}
for the three decoupled oscillators, i.e., the center of mass-photon polaritons, which are created and annihilated by $b^\dagger_j$ and $b_j$ respectively, and have corresponding quantum numbers $n_j$.

Similarly, the Bogoliubov transformation modifies the way in which the spin couples to the CM and photon oscillators. Then, 
\begin{equation}
    \begin{split}
    \mathcal{H}^{cm-ph}_{soc}:=&\mathcal{H}^{cm}_{soc}+\mathcal{H}^{ph}_{soc}, \\
        =&\frac{1}{\sqrt{2}}\vec{a}^\dagger\cdot \vec{\lambda}S_++\text{h.c.},
    \end{split}
\end{equation}
where $\vec{\lambda}^T=\left[\lambda_R \mu_+, -i\lambda_D \mu_-, ig\mu_f, -i\lambda_D \mu_+, \lambda_R\mu_-, -\lambda_D \mu_f\right]$, with $\mu_\pm=\sqrt{\frac{m^*}{4\Bar{\omega}}}(\omega_b\pm2\Bar{\omega})$ and $\mu_f=g\sqrt{\frac{m^*}{\omega_o}},$ takes the form:
\begin{align}
\begin{split}
\mathcal{H}_{soc}^{cm-ph}=&\frac{1}{\sqrt{2}}\vec{b}^\dagger\cdot \mathcal{U}\vec{\lambda} S++\text{h.c.}
\end{split}
\end{align}
Therefore, the new coupling parameters are given by $\tilde{\vec{\lambda}}=\mathcal{U}\vec{\lambda}$. Despite being an analytical calculation, the expression for the coefficients $\tilde{\vec{\lambda}}$ is extensive. However, numerically, it is found that only the frequency of one of the decoupled oscillators, which we will fix as $n_3$, maintains resonance with the spin. This oscillator, in turn, has coefficients $\tilde{\lambda}_3$ and $\tilde{\lambda}_6$ that increase with the cavity coupling, in contrast to what happens with $\tilde{\lambda}_j$ and $\tilde{\lambda}_{j+3}$ for $j=\{1,2\}.$

In summary, with this Bogoliubov transformation in handling the dipolar interaction, only the spin couples to one of the decoupled oscillators with an interaction that increases with the cavity coupling $g$. Consequently, the following maximum quantum numbers ensure a small deviation in the lower energy levels when $g=0.30$ (its maximum analyzed value): $n_1^{max} = n_2^{max} = 3$ for oscillators whose interaction with the spin does not grow with $g$, and $n_3^{max} = 8$ because its coupling depends on $g$. As for the relative quantum numbers, $n_r^{max}$ was set to 2, and $m_r$ ranged from -6 to $n_r$. We assess energy convergence by observing that when we increment our maximum quantum numbers by one, the variation in the three lowest energy levels is less than 0.001 for the most higher coupling scenario examined in this study.

\section{Cavity dependence expected values}\label{ap:cavDep}

\begin{figure*}
    \centering
    \includegraphics[width=0.9\textwidth]{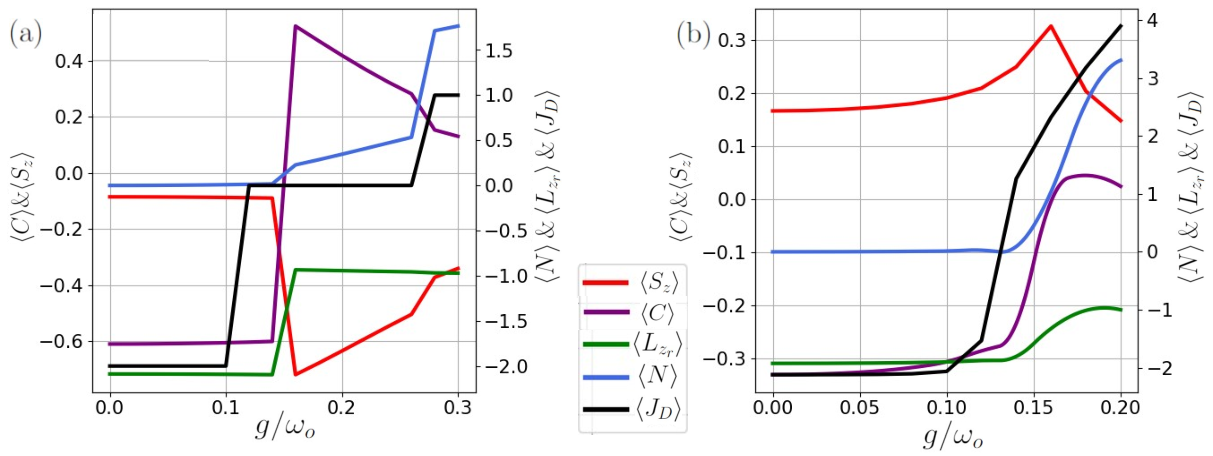}
    \caption{ $\expval{C}, \expval{S_z}, \expval{N}, \expval{L_{z_r}}, \expval{J_D}$ as function of cavity coupling $g$ at $\alpha=3$ (a) with only Dresselhaus $\lambda_D=0.22$ and (b) with both SOCs $\lambda_D=\lambda_R=0.18$.}
    \label{fig:expValG}
\end{figure*}

Despite the experimental challenges in modifying the cavity-matter coupling strength, the results obtained by varying the SOC strength in Fig. \ref{fig:expValG} demonstrate an increase in $g$. Specifically, only Dresselhaus SOC in Fig. \ref{fig:expValG}a exhibits the conserved quantity $J_D$, evident in the staircase-like behavior (black curve) observed in the energy spectrum of the GS. Transitions observed correspond to crossings in the GS energy spectrum, accompanied by simultaneous transitions in other observables. Notably, changes are observed in $\expval{S_z}$ and $\expval{L_z}$, while a distinct transition occurs in $\expval{N}$. These radiation measurements allow for the identification of changes in matter. Furthermore, the cavity is responsible for the transition from $L_{z_r}\sim-2$ to -1 (green curve), indicating a shift in relative density due to the properties of light. The behavior of spin, described by a less pronounced staircase pattern compared to $\lambda_D$, still exhibits transitions in density and a trend towards negative values of $\expval{S_z}$ that tend towards zero (red curve) and uncorrelated spins (purple curve).

In the presence of both SOCs, the absence of a conserved quantity is evident in the smooth behavior of the expected values in Fig. \ref{fig:expValG}b. However, the change from its value at $g=0$ occurs almost simultaneously. In particular, the photon number exhibits an increasing trend as $g$ enhances the contribution of counter-rotating terms between spin and radiation, in contrast to the absence of such contribution in the blue curve of Fig. \ref{fig:expValG}a. In this case, the Rashba SOC contribution leads to positive values of $\expval{S_z}$, but as the cavity introduces interactions with the triplet, a tendency towards cancellation in $\expval{S_z}$ and $\expval{C}$ is observed. The strong entanglement with light states resulting from the increase in $g$ leads to the loss of correlations, making the contribution of triplet states equivalent. Indeed, in the energy spectrum, the onset of variations in different observables is associated with the appearance of degeneracy in the GS due to a greater contribution of light-matter states. Finally, the transition observed in $\expval{L_{z_r}}$ is smoothed out, and this smoothing will have consequences on the charge density, as described elsewhere in this Appendix.

\subsection{Second order perturbation theory for $J_D$ crossings}\label{ap:secondPert}

The observed transitions in the presence of only Dresselhaus SOC, between states with different $J_D$ values, can be analytically explained through second-order perturbation calculations. At $\omega_b=1.9$, there is no degeneracy in the levels under study. In the absence of any coupling ($\lambda_D=\lambda_R=g=0$), the GS is $\ket{A_0}=\ket{n=0, n_{cm}=0, m_{cm}=0, n_r=0, m_r=-2, s=0}$. However, due to interactions, the GS becomes a linear combination of states with $J_D=-2$, with the mentioned state having a higher weight. Conversely, the lowest energy polariton with $J_D=-1$, which is composed by $\ket{B_0}=\ket{n=1, n_{cm}=0, m_{cm}=0, n_r=0, m_r=-1, s=1}$, tends to decrease its energy and become the GS as the couplings increase.

Indeed, perturbative calculations reveal this change in the GS polariton and justify the association between photon number transitions and those observed in spin and angular momentum. This analytical approach takes into account the nearest energy levels, considering that a relatively large energy difference leads to an insignificant contribution. The first mentioned state interacts with the states $\ket{A_1}=\ket{1,0,-1,0, -2, 0}$ due to dipolar interaction, and $\ket{A_2}=\ket{0,0,0,0, -1, 1}$ due to the relative part SOC. Meanwhile, the second state with $n=1$ interacts with the states: $\ket{B_1}=\ket{1,0,-1,0,-1,0}$ due to CM SOC; $\ket{B_2}=\ket{0,1,1,0,-1,1}$ due to dipolar interaction; $\ket{B_3}=\ket{0,0,0,0,-1,0}$ thanks to photonic SOC; and $\ket{B_4}=\ket{1,0,0,0,-2,0}$ due to relative SOC. In fact, entanglement with states having different quantum numbers of spin and angular momentum contributes to the detection of these quantities through light states.

The expression for the energy correction of the first state corresponds to:
\begin{equation}
    \delta E_{A_o}=\frac{|\xi_{dip}(E^{mat}_{A_1}-E^{mat}_{A_0})|^2}{(E^{total}_{A_0}-E^{total}_{A_1})}+\frac{|1.72\xi_{soc}^{rel}(E^{mat}_{A_2}-E^{mat}_{A_0})|^2}{(E^{total}_{A_0}-E^{total}_{A_2})}
\end{equation}
where $E^{mat}_{A_j}=\bar{\omega}(n_{cm}+\frac{1}{2}-\frac{1}{2}(1-\frac{\omega_b}{\bar{\omega}})m_{cm})+\frac{1}{2}\omega_b m_r+\frac{1}{2}\bar{\omega}(\sqrt{m_r^2+\alpha}+1)$ \cite{Quiroga1993SpatialCO} and $E^{total}_{A_j}=E^{mat}_{A_j}+ \Delta_Z s+\Omega n$ for the corresponding quantum numbers of $\ket{A_j}$ and $n_r=0$. On the other hand, the interaction terms are $\xi_{soc}^{rel}=\xi_{soc}^{cm}=\sqrt{2m^*}\lambda_D(\omega_b^2+4\omega_o^2)^{-\frac{1}{4}}$ and $\xi_{dip}=\sqrt{\frac{2}{\omega_o}}g(\omega_b^2+4\omega_o^2)^{-\frac{1}{4}}$. Additionally, the factor of 1.72 arises from the matrix element $ \bra{n_r=0, m_r=-1}r e^{i\phi_r}\ket{n_r=0, m_r=-2}= 1.72 \ell_o$ due to the SOC and was calculated using the analytical solution for the relative degrees of freedom with the $1/r^2$ potential \cite{Quiroga1993SpatialCO}. For the second state, we have:
\begin{equation}
\begin{split}
\delta & E_{B_o}=\frac{|\xi_{soc}^{cm}(E^{mat}_{B_1}-E^{mat}_{B_0})|^2}{(E^{total}_{B_0}-E^{total}_{B_2})}+\frac{|\xi_{dip}(E^{mat}_{B_2}-E^{mat}_{B_0})|^2}{(E^{total}_{B_0}-E^{total}_{B_2})}\\
    &+\frac{|\lambda_D g|^2}{(E^{total}_{B_0}-E^{total}_{B_3})}+\frac{|1.72\xi_{soc}^{rel}(E^{mat}_{B_4}-E^{mat}_{B_0})|^2}{(E^{total}_{B_0}-E^{total}_{B_4})}.
\end{split}
\end{equation}
When calculating the energy difference between the corrected energies, $\Delta E_{AB}=E^{total}_{A_o}-E^{total}_{B_o}+\delta E_{A_o}-\delta E_{B_o}$, we observe that $\lambda_D$ increases this difference, while on the contrary, $g$ through dipolar interaction and photonic SOC reduces $\Delta E_{AB}$. The influence of higher-energy states must be taken into consideration in these calculations in order to accurately diagonalize the Hamiltonian, even though they show evidence for transitions between states with $J_D=-2$ and $J_D=-1$ as the couplings grow. Additionally, the explanation for further polariton transitions in the GS is analogous, as demonstrated previously and in the main text.

\begin{figure}[t]
    \centering    \includegraphics[width=0.45\textwidth]{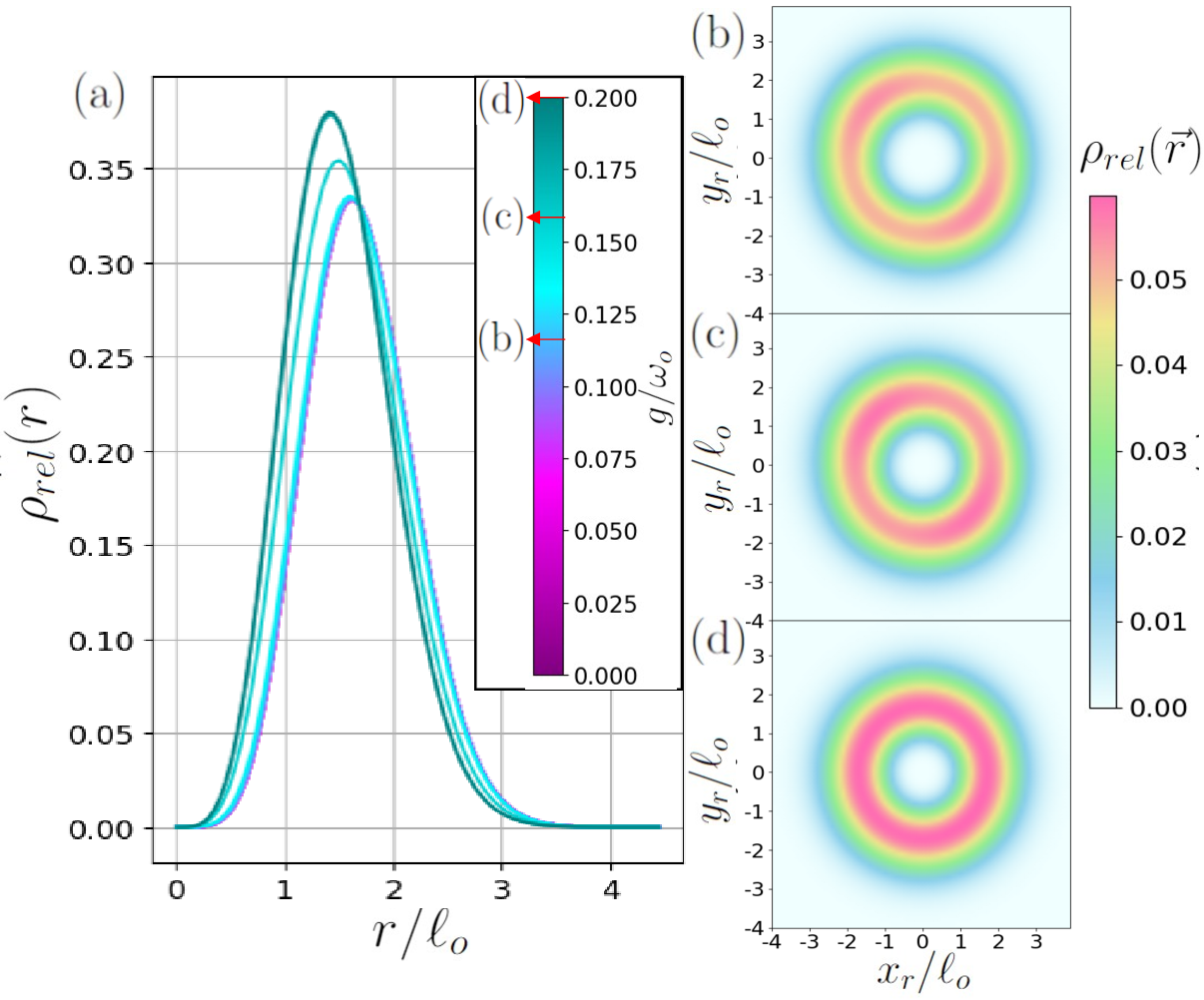}
    \caption{(a) Relative radial density $\rho_r(r):=\int_0^{2\pi} \frac{d\phi}{2\pi}\Tr_{ph, cm, s}(\ket{GS}\bra{GS})$ (tracing over $\phi_r$, CM, photon and spin space) as function of relative radius at fixed $\lambda=\lambda_D=\lambda_R=0,18$ and $\alpha=3$ for different cavity couplings $g$. Respective relative densities $\rho_{rel}(\vec{r}):=\Tr_{ph, cm, s}(\ket{GS}\bra{GS})$ for (b) $g=0.12$, (c) $g=0.16$, (d) $g=0.20$, where colors indicates its spatial probability magnitude.}
    \label{fig:coulombHole}
\end{figure}

\section{Cavity-effective electron-electron interaction}\label{ap:CoulombHole}

In the presence of both SOCs and electron-electron interaction, Luo et al. \cite{LuoNature, Chakraborty} demonstrated that at certain magnetic fields with non-integer expected values, azimuthally asymmetric charge densities are observed. Such densities are shown in Fig. \ref{fig:coulombHole}b, where the density is maximum near $(-1.5, 1.5)$ and $(1.5, -1.5)$, and this pattern is replicated for couplings less than or equal to $g=0.12$. However, as the coupling strength increases along the values where $\langle L_{z_r}\rangle$ smoothly transitions from $-2$ to $-1$, a smooth transition of the charge density occurs as well, as shown in Fig. \ref{fig:coulombHole}a. In this transition, as also evident in Figs. \ref{fig:coulombHole}c and \ref{fig:coulombHole}d, the Coulomb hole reduces smoothly from $r=1.6$ to $r=1.4$. Moreover, the previously observed asymmetry is lost due to the coupling with the states of light and spin. The polariton states that entangle relative space, spin, and radiation start to strongly depend on a series of radiation states that confine the QD to a certain radius, similar to the electron-electron interaction. Therefore, there is measurable evidence of the violation of the Kohn Theorem.

\bibliography{biblio.bib}

\end{document}